\documentclass[10pt]{iopart}
\usepackage{citesort}
\usepackage{cite}
\usepackage{graphicx}
\usepackage{amsfonts}
\usepackage{amssymb}
\usepackage{subfig}
\usepackage{dsfont}
\newcommand{\ket}[1]{\left|{#1}\right\rangle}
\newcommand{\bra}[1]{\left\langle{#1}\right|}

\newcommand{\modu}[1]{\left|{#1}\right|}
\newcommand{\inner}[2]{\left\langle{#1}|{#2}\right\rangle}

\RequirePackage[numbers,sort&compress]{natbib}
\RequirePackage[colorlinks,citecolor=blue,urlcolor=blue,linkcolor=blue]{hyperref}
\begin{document}

\title[Wave packet dynamics of entangled $q$-deformed states]{Wave packet dynamics of entangled $q$-deformed states}
\author{M. Rohith$^{1,2,*}$ \& S. Anupama$^{3}$ \&
        C. Sudheesh$^{3}$}

\address{$^{1}$Quantum Systems Lab, Department of Physics, Government College Malappuram, Malappuram 676 509, University of Calicut, India.\\
        $^{2}$P. G. \& Research Department of Physics, Government Arts and Science College, Kozhikode 673 018, University of Calicut, India. \\
           $^{3}$Department of Physics, Indian Institute of Space Science and Technology,
Thiruvananthapuram, 695 547, India.}
\ead{rohith.manayil@gmail.com}
\vspace{10pt}
\begin{indented}
\item[]February 2024
\end{indented}

\begin{abstract}
This paper explores the wave packet dynamics of a math-type $q$-deformed field interacting with atoms in a Kerr-type nonlinear medium. The primary focus is on the generation and dynamics of entanglement using the $q$-deformed field, with the quantification of entanglement accomplished through the von Neumann entropy. Two distinct initial $q$-deformed states, the $q$-deformed Fock state, and the $q$-deformed coherent state, are investigated. The entanglement dynamics reveal characteristics of periodic, quasi-periodic, and chaotic behavior. Non-deformed initial states display wave packet near revivals and fractional revivals in entanglement dynamics while introducing $q$-deformation eliminates these features. The $q$-deformation significantly influences wave packet revivals and fractional revivals, with even a slight introduction causing their disappearance. For large values of $q$, the entanglement dynamics exhibit a chaotic nature. In the case of a beam splitter-type interaction applied to the initial deformed Fock state, an optimal deformation parameter $q$ is identified, leading to maximum entanglement exceeding the non-deformed scenario.
\end{abstract}

%
%
%
%
\ioptwocol

\section{Introduction}
The concept of quantum physics stems from a profound understanding of operators associated with the observables and their algebras. The deformation to the classical Lie algebras is then developed from a mathematical point of view, in which the value of a deformation parameter characterizes the algebraic operators. The generalized $f$-deformation and $f$-deformed algebras were used to describe many physical situations.
A $q$-deformation is a particular case of $f$-deformed algebra. A $q$-deformation of the quantum harmonic oscillator formalism \cite{Macfarlane_1989} has attained considerable attention through its applications. The $q$-deformed Hamiltonian can be used to describe the energy spectra of certain isotopes \cite{PhysRevC.50.1876}. The deformed Fermi gas model has applications in studying nanomaterials \cite{Algin_2017}. The $q$-deformation was also used in the study of cosmic microwave background radiation \cite{Zeng2017}, the construction of quantum logic gates \cite{Atlantis2014}, for realizing quasibosons \cite{Gavrilik_2011}, etc. The study of several other $q$-deformed systems was also reported in the literature \cite{Raychev_1998,doi:10.1139/cjp-2015-0129,TRISTANT2014276,HATAMI20163469,Hammad_2019}. A comprehensive discourse  on the $q$-deformed generalized Weyl algebra can be found in \cite{mansour2016}.
A recent study on the dynamics of the math-type $q$-deformed harmonic oscillator revealed the signatures of chaos in the system \cite{Pradeep2020}. The realization of the thermostatistics of $q$-deformed algebra can be built on the formalism of $q$-calculus \cite{LAVAGNO2005423}. Nevertheless, the system switches to corresponding non-deformed versions when the deformation vanishes (in the limit $q\rightarrow 1$). It is worth noting that studies on identical nonrelativistic and relativistic objects obeying intermediate statistics are conducted using newly constructed operator algebras \cite{Kuryshkin1988}.

The $q$-deformed states of the electromagnetic field are generally nonclassical states and have potential applications in quantum computation and quantum information protocols. A new kind of $q$-deformed coherent state having $M$-components was developed by extending the concept of Beidenharn $q$-coherent states \cite{doi:10.1142/S0217732396000230}. The nonclassical properties of the even and odd $q$-deformed charge coherent states have been studied \cite{LIU2003210}.   The Hermitian phase difference operators for the two modes of the $q$-deformed electromagnetic field were discussed \cite{Yang1997}. The properties of two-mode $q$-deformed coherent states were also investigated.  The study of nonclassical properties of $q$-deformed noncommutative cat states \cite{PhysRevD.91.044024}, $q$-deformed photon-added nonlinear coherent states \cite{Safaeian_2011}, even and odd $q$-deformed photon-added nonlinear coherent states \cite{Mojaveri2016}, and $q$-deformed superposition states \cite{Anupama_2022} were reported. It is shown that the nonclassical properties of a pair of qubits can be enhanced by introducing the deformation \cite{Metwally2010}. 
 
 Generating entangled states with a large amount of entanglement potential is an important area of research. The entangled states of light are a primary ingredient needed for quantum information processing. The entangled states find applications in the emerging fields of quantum technologies such as quantum cryptography \cite{PhysRevLett.88.187902}, quantum metrology \cite{PhysRevA.94.012101}, superdense coding \cite{PhysRevA.103.022426}, quantum teleportation \cite{PhysRevA.85.054301}, etc. A quantum mechanical beam splitter generates entangled states if one of its input ports contains a nonclassical state \cite{Kim2002,Rohith2016b}.   Recently, the entanglement in the deformed states was investigated \cite{BERRADA2011628, KHALIL2021103720,PhysRevResearch.2.043305}. It was shown that the beam splitter action of the time-evolved states in a nonlinear medium could produce an arbitrarily large amount of entanglement  \cite{vanEnk2003,Sudheesh2006, Rohith2016b}. The dynamics of entangled two-mode states in a nonlinear medium may also show near revivals \cite{Sudheesh2006}.   Nevertheless, a detailed study on the entanglement properties of $q$-deformed fields propagating in a nonlinear medium has not been reported.  
 
 This paper aims to study the entanglement dynamics of the math-type $q$-deformed fields propagating in a Kerr-like atomic nonlinear medium. We also examine the role of field deformation on the extent to which the revivals occur in the entanglement dynamics. The interaction between the $q$-deformed field mode and atomic mode is assumed to be a beam splitter type of interaction. We study the entanglement dynamics of the initial $q$-deformed Fock state and $q$-deformed coherent states propagating in the nonlinear medium. The rest of the manuscript is organized as follows. Section \ref{sec2} discusses the  Hamiltonian model used for investigating the entanglement dynamics of initial $q$-deformed states.  The formalism used for quantifying the entanglement as a function of time for an initial $q$-deformed state is given in Sec.~\ref{sec3}. Section {\ref{sec4}} describes our numerical simulation results for initial $q$-deformed Fock states and $q$-deformed coherent states propagating in Kerr-like atomic nonlinear medium. Our main results of this paper are presented in Sec. \ref{sec5}.

\section{Model}\label{sec2}
Consider the propagation of a single-mode math-type $q$-deformed field through a Kerr-like atomic nonlinear medium.
The Kerr effect is a nonlinear optical phenomenon that occurs when light propagates through media such as crystals or glasses. This effect is characterized by a change in the refractive index induced by electric fields, which is proportional to the square of the electric field strength. It has been demonstrated that Kerr nonlinearity can effectively generate entanglement within arbitrarily short time durations during standard nonlinear optical interactions, which are subsequently followed by interactions with a beam splitter \cite{vanEnk2003,Rohith2016b}.
  We assume a beam splitter kind of interaction between the $q$-deformed field mode and the atomic mode of the system. The total Hamiltonian representing the interaction is written as
\begin{equation}
    H_{tot}=H_q+H_{a}+H_{int}, 
\end{equation}
where $H_q$ represents the Hamiltonian of the math-type $q$-deformed field, $H_{a}$ represents the Hamiltonian of the atomic nonlinear medium, and $H_{int}$ is the Hamiltonian representing the interaction between the deformed field and the atoms of the nonlinear medium. The math-type $q$-deformation of the field is governed by the Hamiltonian \cite{Eremin2006}
\begin{equation}
     H_{q} = \frac{1}{2} (AA^{\dagger} + A^{\dagger}A),\label{H_q}
\end{equation}
where  $q$ ranges from $0$ to $1$, $A$ and $A^{\dagger}$ are the annihilation and creation operators of the deformed field, which obey the deformed commutation relation
\begin{equation}
    A A^{\dagger} - q^{2} A^{\dagger} A = \mathds{1}.
\end{equation}
The action of $A$ and $A^{\dagger}$ on the deformed Fock state $\ket{n}_{q}$ is defined by \cite{PhysRevD.87.084033},
\begin{eqnarray}
     A\ket{n}_{q} &=& \sqrt{[n]}\ket{n-1}_{q}, \\ 
     A^{\dagger}\ket{n}_{q} &=& \sqrt{[n+1]}\ket{n+1}_{q}, 
\end{eqnarray}
where
\begin{equation}\label{box_n}
    [n] = \frac{1-q^{2n}}{1-q^{2}}.
\end{equation}
Here and in the rest of the manuscript, we choose $\hbar=1$. In the limit $q \rightarrow 1$ ($q=1$ corresponds to the non-deformed case), $[n] \rightarrow n$ and the Hamiltonian (\ref{H_q}) reduces to the usual non-deformed harmonic oscillator Hamiltonian
\begin{equation}
    H = \frac{1}{2} (aa^{\dagger} + a^{\dagger}a),
\end{equation}
where $a$ and $a^{\dagger}$ are the non-deformed annihilation and creation operators, respectively, which obey the commutation relation
\begin{equation}
    a a^{\dagger} - a^{\dagger} a = \mathds{1}.
\end{equation}
The Hamiltonian of the nonlinear atomic medium is taken as
\begin{equation}
    H_{a} = \omega \left(b^{\dagger} b+1/2\right)+\chi {b^{\dagger}}^{2}b^{2},
\end{equation}
where $b$ and $b^{\dagger}$ are the atomic ladder operators, $\omega$ is the natural frequency of the atomic field,  and $\chi$ is the nonlinearity parameter. 
\begin{figure*}[h]
\includegraphics[scale=0.7]{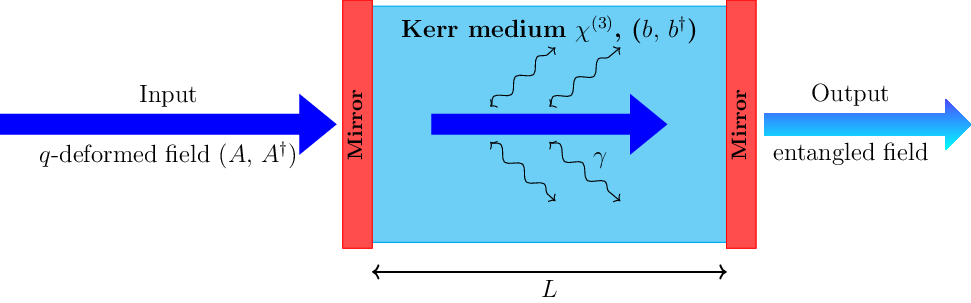}
\centering
\caption{Schematic diagram illustrating the interaction between a \(q\)-deformed field and a Kerr nonlinear medium. The propagation time of the initial state through the medium is adjustable by varying the length \(L\) of the Kerr medium.}
\label{schematic}
\end{figure*}

The interaction between the deformed field and the atomic medium is represented by the interaction Hamiltonian
\begin{equation}\label{Hint}
    H_{int}= \gamma (A^{\dagger}b+Ab^{\dagger})
\end{equation}
where $\gamma$ is the coupling strength between the field and atomic modes. The time-evolved state of the system at any instant can be written as
\begin{equation}
    \ket{\psi(t)} = \exp\left[-i\,H_{tot}t\right] \ket{\psi(0)}.
\end{equation}
The system's dynamics explicitly depend on the initial states $\ket{\psi(0)}$ taken. Throughout this study, we consider the atom in the ground state $\ket{0}$. Therefore, the initial unentangled direct product states can be written as $\ket{\phi}_q\otimes\ket{0}$, where the $\ket{\phi}_q$ is the deformed field mode. We examine the entanglement dynamics for the field $\ket{\phi}_q$, which is initially prepared in the deformed Fock state $
\ket{N}_{q}$ and in the deformed coherent state $\ket{\alpha}_{q}$. In the subsequent section. We describe the quantification of entanglement shown by the system in terms of von Neumann entropy. 

\section{Entanglement dynamics} \label{sec3}
We use the von Neumann entropy of the subsystem $S_i$ as the measure of entanglement, where the suffix $i$ stands for either $q$ or $b$, depending upon the subsystem considered. If $\rho_i$ represents the time-dependent reduced  density matrix for the subsystem, the von Neumann entropy is defined as
\begin{equation}
    S_i=-{\rm Tr}_i\left[\rho_i\left(t\right) \ln \rho_i\left(t\right)\right].\label{vne}
\end{equation}
As the total particle number $\mathcal{N}_{tot}=A^\dagger A+b^\dagger b$, is conserved during the interaction (One can show that $\left[\mathcal{N}_{tot}, H_{tot}\right] = 0$), we choose the basis states as $\ket{N-m}_q\otimes \ket{m}_a\equiv \ket{(N-m)_q;m}$, where $N$ is the eigenvalue of $\mathcal{N}_{tot}$. Here $N$ runs from $0$ to $\infty$ and $m$ runs from $0$ to $N$. One can see that $\bra{(N-m)_q;m}H_{tot}\ket{(N^\prime-m^\prime)_q; m^\prime}=0$ for $N\neq N^\prime$. Hence for a particular $N$, the total Hamiltonian $H_{tot}$ can be diagonalized in the space spanned by the set \{$\ket{(N-m)_q;m}$\} with $m=0,\,1,\,2,\, \dots,\, N$. Let the eigenvalues and eigenvectors of $H_{tot}$ be $\lambda_{Nj}$ and $\ket{\psi_{Nj}}_q$, respectively. Here the index $j$ designate the eigenvectors in each block of the Hamiltonian $H_{tot}$ for a particular $N$, that is, $j=0,\,1,\,2,\,\dots,\,N$. The eigenvectors $\ket{\psi_{Nj}}_q$ can be expanded in the basis \{${\ket{(N-m)_q;m}}$\} as
\begin{equation}
    \ket{\psi_{Nj}}_{q} = \sum\limits_{m=0}^{N} C^{Nj}_{m} \ket{(N-m)_q;m},
\end{equation}
where $C^{Nj}_{m} = \inner{\left(N-m\right)_q;\, m}{\psi_{Nj}}_q$ are the $q$ dependent expansion coefficients. An initial state of the system evolves in time as
\begin{eqnarray}
    \ket{\psi(t)} &&= \exp \left[-i\,H_{tot}\, t\right] \ket{\psi(0)}\\
    &&= \sum \limits_{N=0}^{\infty} \sum \limits_{j=0}^{N} e^{-i\lambda_{Nj}\,t} {_q}\left<{\psi_{Nj}}\vert \psi(0)\right> \ket{\psi_{Nj}}_q.
\end{eqnarray}
The time-evolved density matrix of the total system is calculated as
\begin{eqnarray}
    \rho_{tot}(t) =&& \sum\limits_{N=0}^{\infty}\sum\limits_{j=0}^{N}\sum\limits_{N^{'}=0}^{\infty}\sum\limits_{j^{'}=0}^{N^{'}} e^{-i(\lambda_{Nj}-\lambda_{N^{'}j^{'}})\,t} {_q}\left<{\psi_{Nj}}\vert \psi(0)\right>\nonumber\\
    && \times \left<{\psi(0)}\vert {\psi_{N^{'}j^{'}}}\right>_q \ket{\psi_{Nj}}_q {_q}\bra{\psi_{N^{'}j^{'}}}.\label{denistyfull}
\end{eqnarray}
We use Eq.~(\ref{denistyfull}) to calculate the system's bipartite entanglement as a function of time.

\section{Results and discussion}\label{sec4}
In this study, we quantify the bipartite entanglement between the system's deformed field mode and the atomic mode. Our analysis focuses on two distinct initial states of the field: the initial deformed Fock state and the initial deformed coherent state.

\subsection{Field initially in the deformed Fock state $\ket{N}_q$}
Here, we assume the initial state of the field to be the deformed Fock state $\ket{N}_{q}$, while the atom is in the ground state $\ket{0}$. The resultant time-evolved reduced density matrix of the system, denoted as $\rho_{k}(t)$ ($k=q$ denotes the deformed field mode and $k=a$ denotes the atomic mode), can be expressed as follows:
\begin{eqnarray}
    \rho_{q}(t) =&& \sum\limits_{m=0}^{N}\sum\limits_{j=0}^{N}\sum\limits_{j^{'}=0}^{N} \exp\left[-i(\lambda_{Nj}-\lambda_{N j^{'}})\,t\right] {C_{0}^{Nj}}^\ast C_{0}^{Nj^{'}}\nonumber\\
    && \times C_{m}^{Nj} {C_{m}^{Nj^{'}}}^\ast \ket{N-m}_{q}{}_{q}\bra{N-m}\label{Fockrhosubq}
\end{eqnarray}
and
\begin{eqnarray}
    \rho_{a}(t) =&& \sum\limits_{m=0}^{N}\sum\limits_{j=0}^{N}\sum\limits_{j^{'}=0}^{N} \exp\left[-i(\lambda_{Nj}-\lambda_{N j^{'}})\,t\right] {C_{0}^{Nj}}^\ast C_{0}^{Nj^{'}} \nonumber\\
    && \times C_{N-m}^{Nj} C_{N-m}^{{N j^{'}}^\ast}\ket{N-m}_{a}{}_{a}\bra{N-m},\label{Fockrhosubb}
\end{eqnarray}
where the coefficient $C_{0}^{Nj} = \langle N_{q};0\vert\psi_{Nj}\rangle$.

\begin{figure}[h]
\centering	
\includegraphics[scale=0.5]{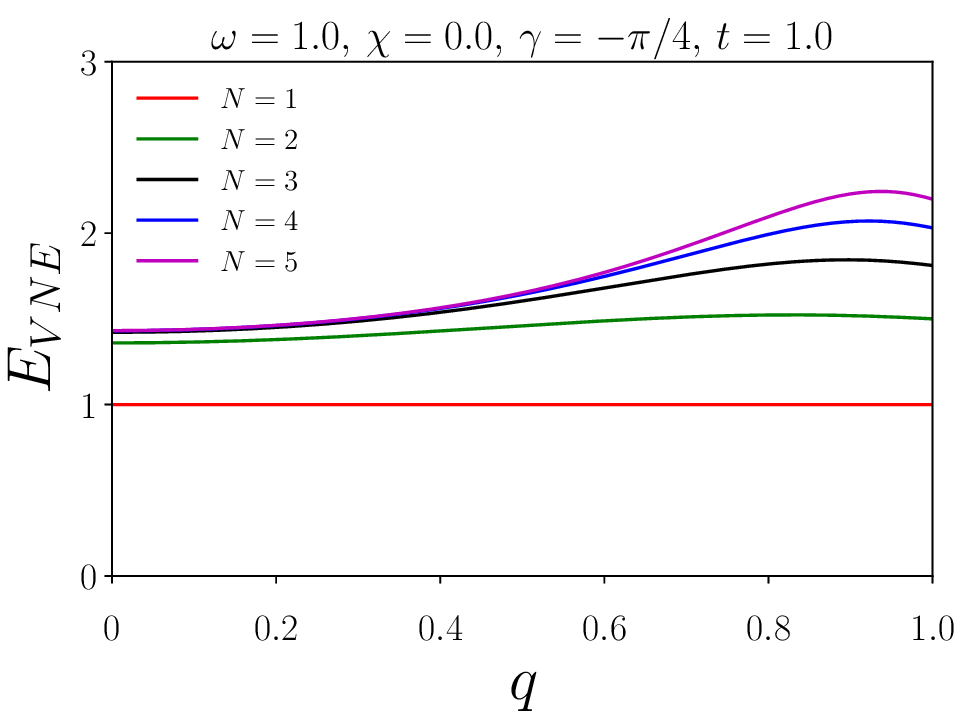}
\caption{Variation of von Neumann entropy with $q$ across various values of $N$ at time $t=1.0$.}\label{EntQvariation}
\end{figure}
The system's entanglement is subsequently quantified through the numerical evaluation of von Neumann entropy, utilizing Eqs.~(\ref{vne}), (\ref{Fockrhosubq}), and (\ref{Fockrhosubb}). We have set the parameter values to $\omega=1$, $\chi=0$, $\gamma= -\pi/4$, and $t=1$. When $\gamma=-\pi/4$, the interaction Hamiltonian (10) transforms into the unitary operator representing the deformed scenario's beam splitter. The variation of von Neumann entropy with the deformation parameter $q$ is depicted in Fig.~\ref{EntQvariation} for different values of $N$. When $q$ equals $1$, the entanglement value corresponds to the non-deformed scenario, wherein the field is in the Fock state $\ket{N}$, and the atom is in the ground state $\ket{0}$. A considerable variation can be seen for $N=1$ from the other $N$ values. When $N=1$, the plot illustrates that the change in system entanglement, compared to the non-deformed case, is negligible. When $N>1$, the figure indicates the existence of an optimal deformation parameter value. At this optimal point, the entanglement reaches its maximum, surpassing the value associated with the non-deformed case. For instance, with $N=5$, the maximum entanglement is achieved at $q=0.937$, resulting in a value of $2.243$, which exceeds the entanglement of the corresponding non-deformed case ($E_{VNE}=2.198$). Consequently, this indicates that the field's distortion can produce a greater degree of entanglement than its non-deformed counterpart. As the value of $N$ increases, there is an observed shift in the optimal deformation parameter, indicating maximum entanglement, towards $q=1$.

\begin{figure}[h]
    \centering
	\subfloat[]{\includegraphics[width=0.85\linewidth]{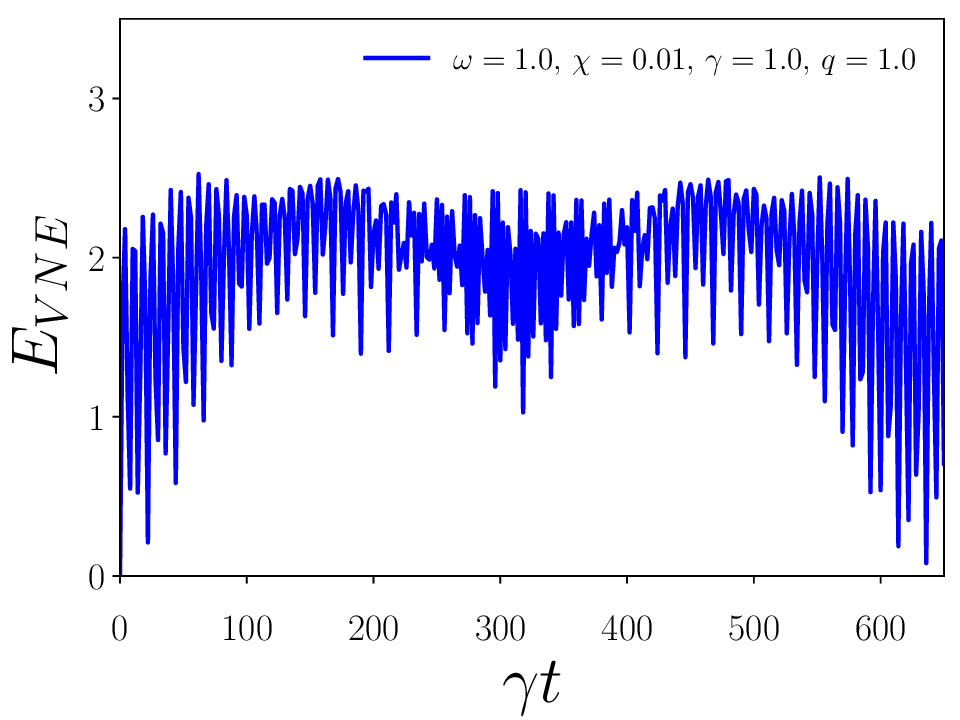}\label{fig2a}}\\
	\subfloat[]{\includegraphics[width=0.85\linewidth]{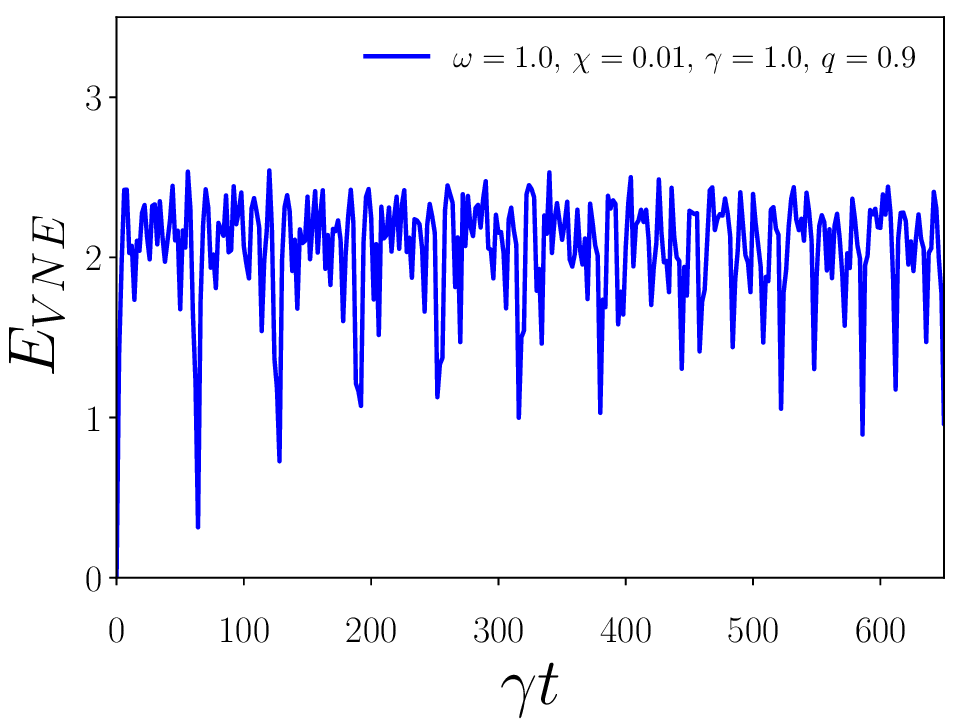}\label{fig2b}} \\
	\subfloat[]{\includegraphics[width=0.85\linewidth]{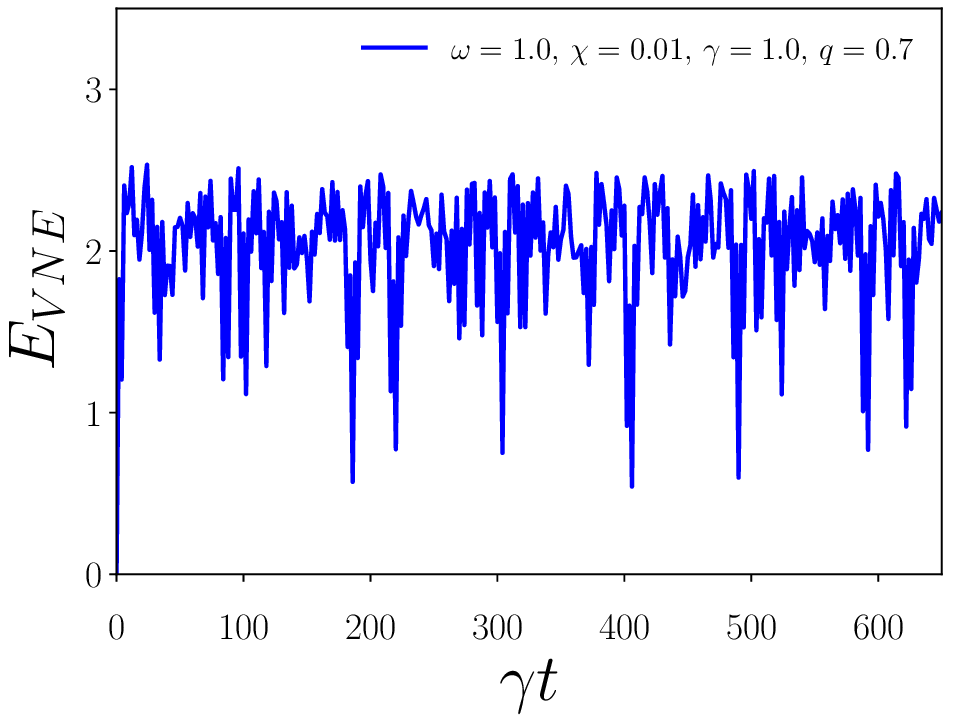}\label{fig2c}}
    \caption{Variation of von Neumann entropy with $\gamma t$ for (a) non-deformed, (b) $q=0.9$, and (c) $q=0.7$ for the initial state $\ket{5}_q\otimes\ket{0}_a$.}
   \label{fig:VNE_for_state_5_0}
\end{figure}

When an initial wave packet propagates within the Kerr-like atomic medium, it experiences dispersion. The revival phenomenon denotes the reconstitution of the dispersed wave packet to its initial form. The two-subpacket fractional revival time marks the moment when the wave packet becomes a superposition of two identical copies of the initial wave packet. We examine the dynamics of entanglement in the system for different deformation values. Since the system's weak non-linearity encourages the occurrence of near revival phenomena, we have chosen a smaller nonlinearity parameter of $\chi/\gamma = 0.01$ in all our computations. The values for the remaining parameters are set as follows: $\omega = 1.0$, and $\gamma = 1.0$.  The periodic energy transfer between the field and atomic modes may result in quasi-revivals occurring at times roughly corresponding to integer multiples of $2\pi/\chi$. Figure~\ref{fig:VNE_for_state_5_0} illustrates a comparison of the von Neumann entropy of the system under various deformations, specifically for the initial state $\ket{5}_q\otimes\ket{0}_a$.  Figure~\ref{fig2a} depicts the non-deformed scenario, where the entropy periodically returns to values near zero at   approximately equal to $\gamma t\approx 2\pi\times 10^{2}$. This observation signifies the existence of a quasi-revival phenomenon within the system \cite{Sudheesh2006}. 

Between time $t=0$ and $t=2\pi/\chi$, there are rapid fluctuations in the entropy value. A relative minimum value of entropy, observed in the plot around $\gamma t\approx \pi/\chi$, signifies the occurrence of two-subpacket fractional revivals.  Figure~\ref{fig2b} illustrates the entropy measured in the system for a slight deformation, say $q=0.9$. The near revivals are less pronounced than in the case of the non-deformed system. Even with a slight deformation, the two-subpacket fractional revival  is erased from the system. The degree of both revivals and fractional revivals undergoes a significant reduction when the deformation parameter is set to $q=0.7$ [see Fig.~\ref{fig2c}]. With $q=0.7$, the entropy value oscillates so rapidly that revivals and fractional revivals in the medium are not   evident. 

\begin{figure}[h]
    \centering
	\subfloat[]{\includegraphics[width=0.85\linewidth]{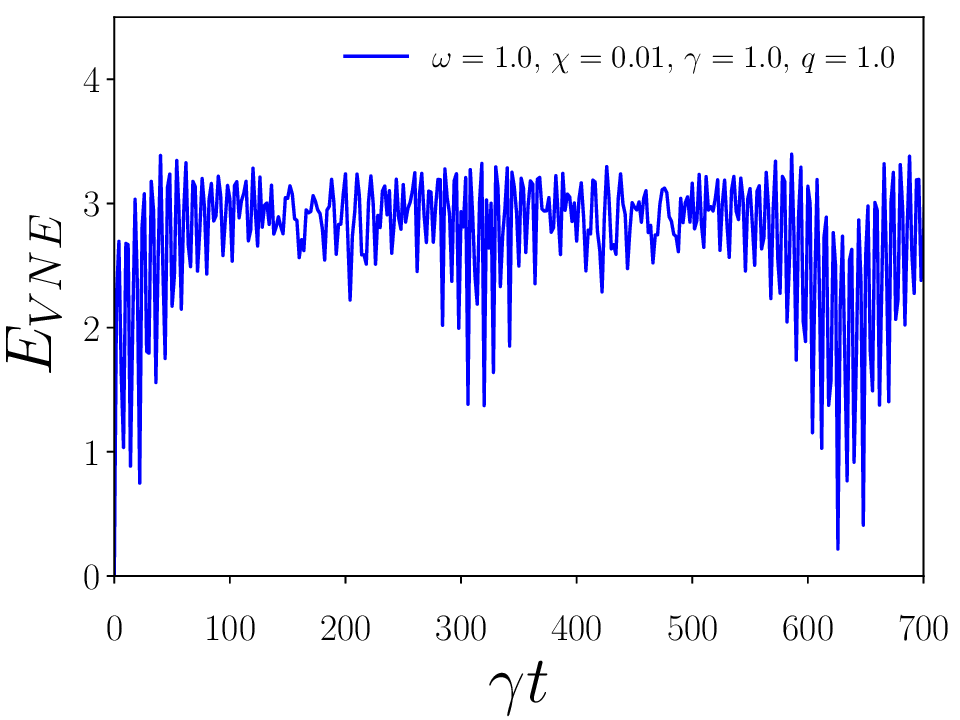}\label{fig3a}}\\
	\subfloat[]{\includegraphics[width=0.85\linewidth]{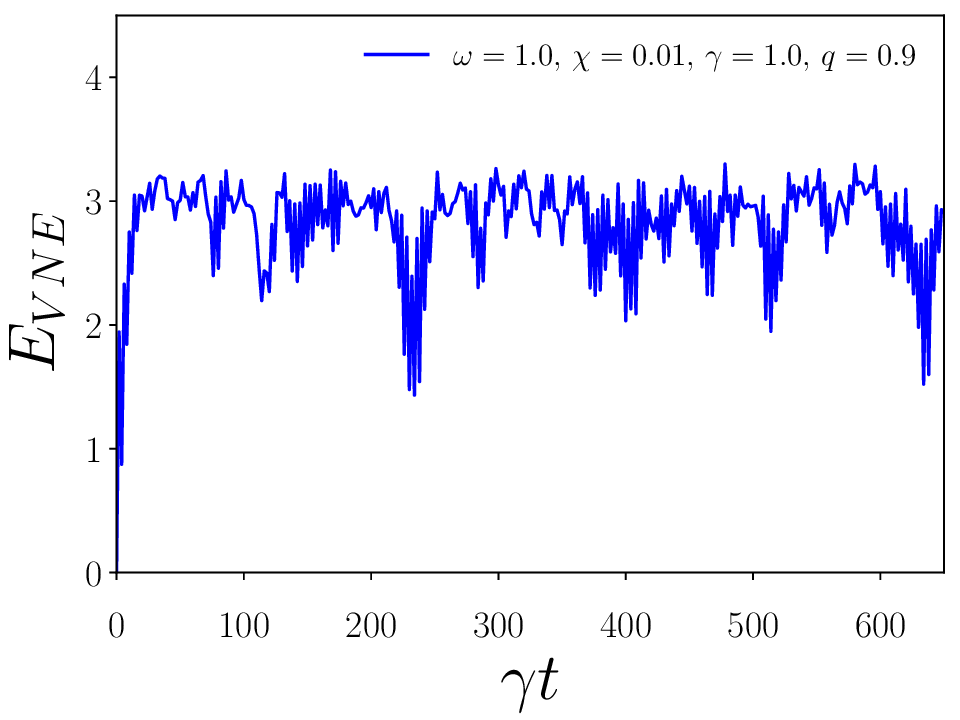}\label{fig3b}} \\
	\subfloat[]{\includegraphics[width=0.85\linewidth]{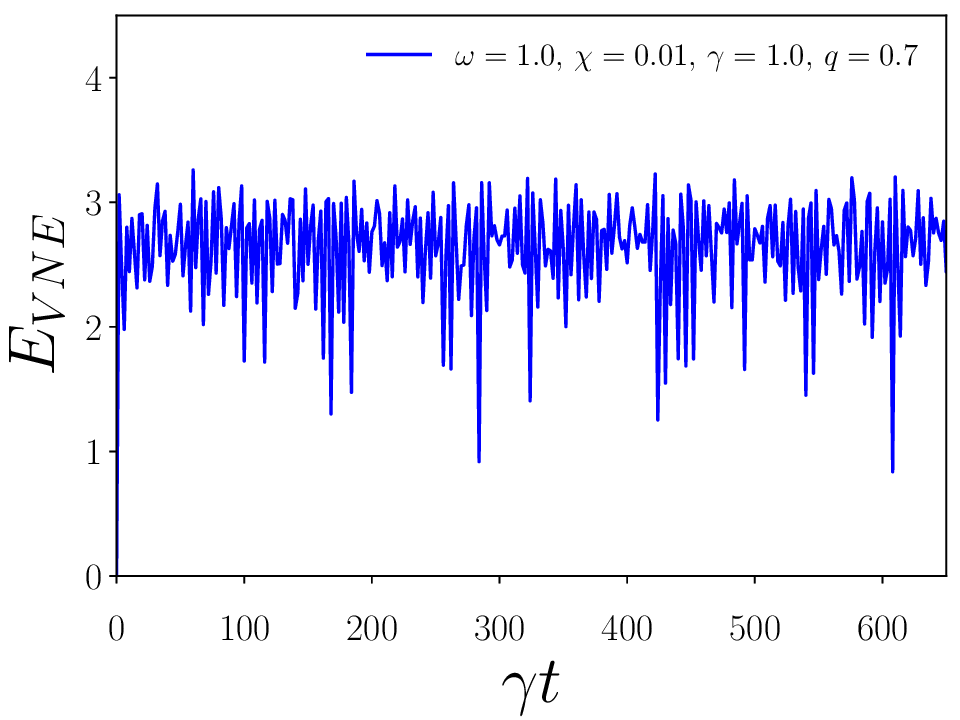}\label{fig3c}}
    \caption{Variation of von Neumann entropy with $\gamma t$ for (a) non-deformed, (b) $q=0.9$, and (c) $q=0.7$, for the initial state $\ket{10}_q\otimes\ket{0}_a$.}
   \label{fig:VNE_for_state_10_0}
\end{figure}

We conducted calculations for various initial Fock states and obtained similar results. Figure~\ref{fig:VNE_for_state_10_0} illustrates the temporal evolution of entropy for an initial Fock state $\ket{10}_q\otimes\ket{0}_a$. In Fig.~\ref{fig3a}, the entropy of the non-deformed initial state $\ket{10}\otimes\ket{0}_a$ is depicted, revealing notable near-revivals in entropy dynamics. The local minima observed in entropy values between time $t=0$ and $t=2\pi/\chi$ signify fractional revivals. It is evident that, even with a slight deviation from the non-deformed case (i.e., for $q=0.9$), the revivals vanish from the system (See Fig.~\ref{fig3b}). Figure~\ref{fig3c} indicates that, with increasing deformation, the revivals gradually diminish, exhibiting a faster deviation than those observed for the initial state $\ket{5}_q\otimes\ket{0}_a$. For a given initial Fock state, despite the small nonlinearity, both revivals and fractional revivals gradually disappear with an increase in the field's deformation. The calculations mentioned above are reiterated for the high nonlinearity of the system, revealing that the observed characteristics described earlier are duplicated in this scenario as well.

\subsection{Field initially in the deformed coherent state}
Next, we consider the scenario wherein the field starts in the deformed coherent state $\ket{\alpha}_q$, and the atom resides in its ground state $\ket{0}$, thereby representing the initial state as $\ket{\psi \left(0\right)}=\ket{\alpha}_q \otimes \ket{0}_a$. The $q$-deformed coherent state, denoted as $\ket{\alpha}_q$, is expressed by the equation
\begin{equation}
\ket{\alpha}_q= e_q^{-\modu{\alpha_q}^2/2}\sum_{n=0}^{\infty} \frac{\alpha_q^n}{\sqrt{\left[n\right]!}} \ket{n}_q,
\end{equation}
where the $q$-deformed exponential function is defined as
\begin{equation}
e_q^{(\bullet)}=\sum_{n=0}^{\infty} \frac{(\bullet)^n}{\left[n\right]!}.
\end{equation}
Upon substituting the initial state $\ket{\psi \left(0\right)}$ into Eq.~(\ref{denistyfull}), we obtain the reduced density matrix $\rho_{k}$ for the subsystems as
\begin{eqnarray}
    \rho_{q}(t) &&= e_q^{-\modu{\alpha_q}^2}\sum\limits_{m=0}^{\infty}\sum\limits_{N=0}^{\infty}\sum\limits_{j=0}^{N}\sum\limits_{N^{'}=0}^{\infty}\sum\limits_{j^{'}=0}^{N^{'}}\frac{\alpha^{N}(\alpha^{*})^{N^{'}}}{\sqrt{[N]![N^{'}]!}}\nonumber\\
    &&\times \exp\left[-i(\lambda_{Nj}-\lambda_{N^{'} j^{'}})\,t\right] {C_{0}^{Nj}}^\ast C_{0}^{N^{'}j^{'}}\nonumber\\
    &&  \times   C_{m}^{Nj} C_{m}^{{N^{'}j^{'}}^\ast}\ket{N-m}_{q}{}_{q}\bra{N^{'}-m}.\label{reducedCSdensity1}
\end{eqnarray}
and
\begin{eqnarray}
    \rho_{a}(t) &&= e_q^{-\modu{\alpha_q}^2}\sum\limits_{m=0}^{\infty} \sum\limits_{N=0}^{\infty}\sum\limits_{j=0}^{N}\sum\limits_{N^{'}=0}^{\infty}\sum\limits_{j^{'}=0}^{N^{'}}\frac{\alpha^{N}(\alpha^{*})^{N^{'}}}{\sqrt{[N]![N^{'}]!}}\nonumber\\
    &&\times \exp\left[-i(\lambda_{Nj}-\lambda_{N^{'} j^{'}})\,t\right]{C_{0}^{Nj}}^\ast C_{0}^{N^{'}j^{'}}\nonumber\\
    && \times  C_{N-m}^{Nj} C_{N-m}^{{N^{'}j^{'}}^\ast}\ket{N-m}_{a}{}_{a}\bra{N^{'}-m}. \label{reducedCSdensity2}
\end{eqnarray}

\begin{figure}[ht]
    \centering
	\subfloat[]{\includegraphics[width=0.85\linewidth]{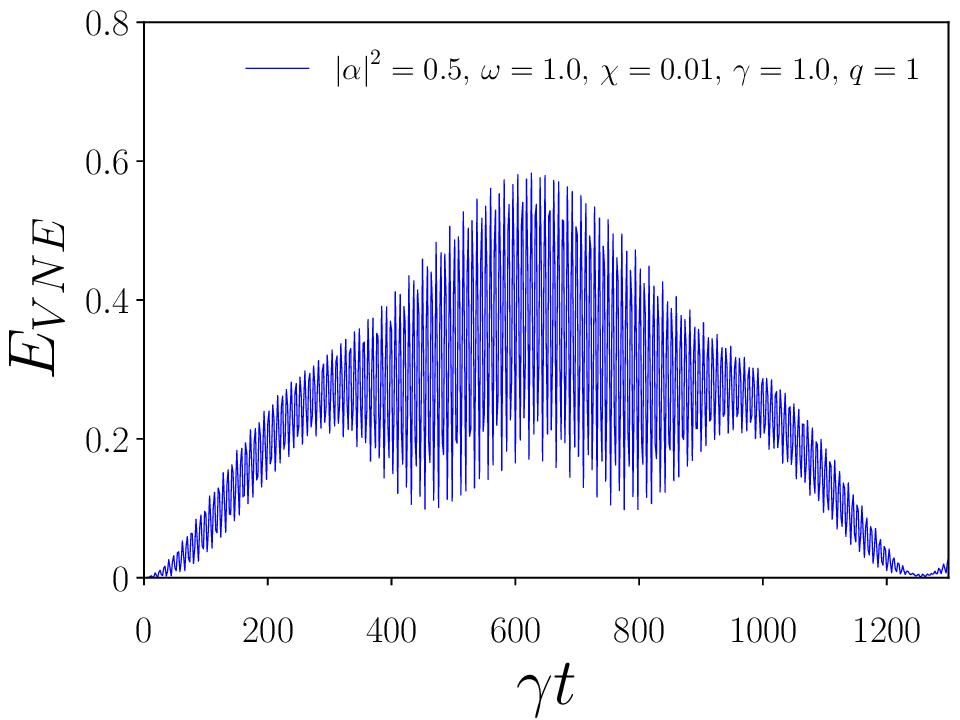}\label{fig4a}}\\
	\subfloat[]{\includegraphics[width=0.85\linewidth]{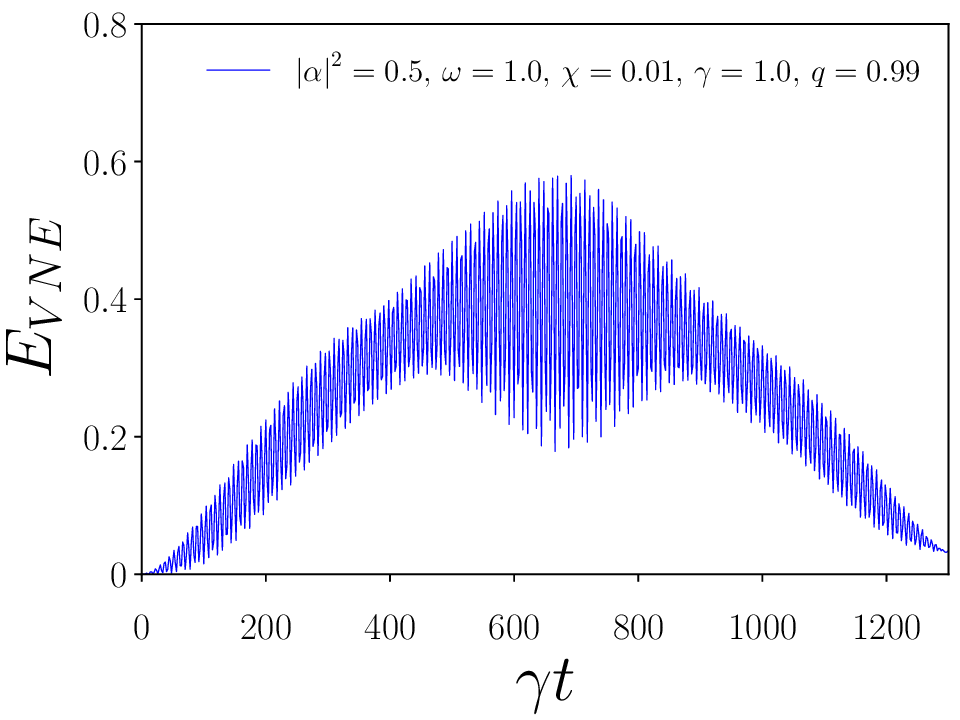}\label{fig4b}} \\
	\subfloat[]{\includegraphics[width=0.85\linewidth]{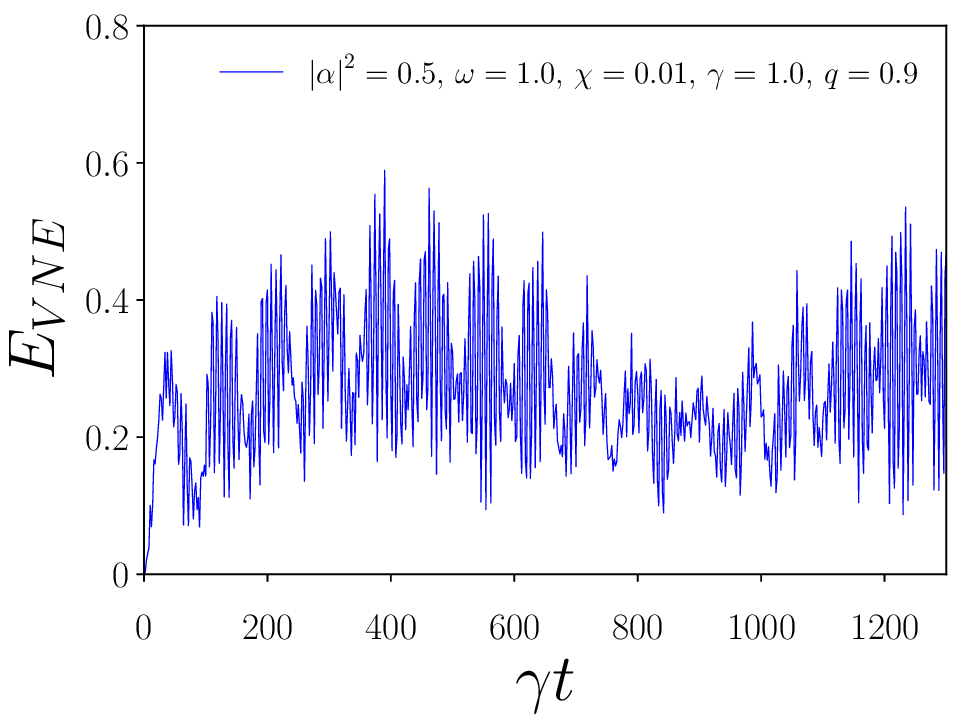}\label{fig4c}}
    \caption{Evolution of von Neumann entropy with $\gamma t$ for (a) undeformed, (b) $q=0.99$, and (c) $q=0.9$, considering the initial state $\ket{\alpha}_q \otimes \ket{0}_a$ with $\vert \alpha\vert^{2} = 0.5$.}
   \label{fig:VNE_for_state_alpha_0_mod05}
\end{figure}

The von Neumann entropy of the subsystems has been computed over time, utilizing Eqs.~(\ref{vne}), (\ref{reducedCSdensity1}), and (\ref{reducedCSdensity2}). In this case, we have selected the parameters $\omega=1$, $\chi=0.01$, and $\gamma=1$. The time evolution of the von Neumann entropy for the initial state $\ket{\alpha}_{q}\otimes \ket{0}$ with $\vert \alpha \vert^{2}=0.5$ and varying values of the deformation parameter $q$ is illustrated in Fig.~(\ref{fig:VNE_for_state_alpha_0_mod05}). The degree of entanglement predominantly relies on the parameter $\alpha$, and its value fluctuates over time. In the case of no deformation, it is evident that the von Neumann entropy consistently reverts to its initial value of zero at regular time intervals, demonstrating the phenomenon of wave packet revival. The system experiences a revival period of $4\pi/\chi$. Even a minor departure from the pure case in the deformation parameter $q$ not only eliminates the wave packet revival but also becomes evident in the entanglement dynamics of the system, as depicted in Fig.~\ref{fig4b}. 

It has been demonstrated that the time series of the anticipated values of dynamic variables, such as observables related to position and momentum, display periodic, quasi-periodic, or chaotic behavior contingent upon the deformation parameter $q$ \cite{Pradeep2020}. Similarly, in this scenario, we have distinctly observed the indications of periodic, quasi-periodic, or chaotic behavior in the entanglement dynamics of the system. For the non-deformed scenario with $q=1$ as depicted in Fig.~\ref{fig4a}, there is an evident periodic behavior in the entanglement dynamics of the system. With slight deformations, the entanglement dynamics exhibit quasi-periodic behavior, as illustrated in Fig.~\ref{fig4b}. Under substantial deformation, the entanglement value undergoes chaotic fluctuations, as depicted in Fig.~\ref{fig4c}. The numerical analysis described above was replicated for the initial deformed coherent state, considering various values of $\modu{\alpha}$, and consistent results were obtained.

\section{Conclusion}\label{sec5}
In conclusion, this paper delved into the wave packet dynamics of a math-type $q$-deformed field interacting with atoms in a Kerr-type nonlinear medium. Our investigation focused on the generation and dynamics of entanglement using the  $q$-deformed field. A schematic diagram illustrating the interaction of a \(q\)-deformed input field with an atomic Kerr nonlinear medium placed between two mirrors is presented in Fig.~\ref{schematic}. In the figure, we have explicitly retained \(\chi^{(3)}\) to emphasize its role as the third-order nonlinear susceptibility of the Kerr medium.
To quantify the entanglement, we employed the von Neumann entropy. The entanglement dynamics were analyzed for two distinct initial $q$-deformed states: the $q$-deformed Fock state and the $q$-deformed coherent state. We observed that the entanglement dynamics display indications of periodic, quasi-periodic, and chaotic behavior in their evolution.
Non-deformed initial states demonstrate wave packet near revivals and fractional revivals in the entanglement dynamics, while the introduction of  $q$-deformation eliminates these near revivals from the system. The $q$-deformation of the field markedly influences the wave packet revivals and fractional revivals. Even a slight introduction of deformation causes their disappearance. The entanglement dynamics exhibit a chaotic nature for large values of $q$. In the case of a beam splitter-type interaction applied to the initial deformed Fock state, we observed that there exists an optimal deformation parameter $q$. At this optimum, the system exhibits maximum entanglement, surpassing the value associated with the non-deformed scenario. All the results  demonstrate that the entanglement in the system is significantly influenced by both the deformation of the system and the field strength. Therefore, the introduction of deformation introduces an additional degree of freedom, denoted as $q$, enabling control over the entanglement. As deformed states can more closely emulate the states of real systems compared to ideal non-deformed cases, creating $q$-deformed entangled states and measuring entanglement over time could prove beneficial for applications in quantum information processing.

\section*{Data Availability Statement} 
No data is used to produce any result in the paper. All the ﬁgures in the manuscript are produced using the equations derived in the manuscript.

\ack
The conception of the current concept originated from C Sudheesh, while S. Anupama performed the analytical calculations. M Rohith conducted the numerical analysis for the problem, and S. Anupama, with M. Rohith's assistance, prepared the manuscript. C Sudheesh oversaw the entire project. M. Rohith expresses gratitude for the financial support provided for this research by the Science and Engineering Research Board, Government of India, through the State University Research Excellence (SERB-SURE)  scheme, with reference number SUR/2022/003354.
\bibliography{reference}
\end{document}